\documentclass[fleqn]{SCGE2020}
\usepackage{hyperref}  
\newcommand{\msun}{{\rm M}_\odot}

\newcommand{\beq}{\begin{equation}}
\newcommand{\eeq}{\end{equation}}
\newcommand\lsim{\mathrel{\rlap{\lower4pt\hbox{\hskip0pt$\sim$}}
        \raise1pt\hbox{$<$}}}
\newcommand\gsim{\mathrel{\rlap{\lower4pt\hbox{\hskip0pt$\sim$}}
        \raise1pt\hbox{$>$}}}

\begin{document}
 
\ensubject{subject}


\ArticleType{Article}%
\Year{2026}
\Month{March}
\Vol{69}
\No{3}
\DOI{10.1007/s11433-025-2850-5}
\ArtNo{230412}
\ReceiveDate{September 5, 2025}
\AcceptDate{November 7, 2023}
\OnlineDate{January 4, 2016}

\title{Revealing the origin of supermassive black holes with Taiji-TianQin network}{Revealing the origin of supermassive black holes with Taiji-TianQin network}

\author[1,2]{Ping Shen}{}
\author[1,2,3,4]{Wen-Biao Han}{wbhan@shao.ac.cn}
\author[5]{Wen-Xin Zhong}{}

\AuthorMark{P. Shen}%
\AuthorCitation{P. Shen, W.-B. Han, and W.-X. Zhong}

\address[1]{Shanghai Astronomical Observatory, Chinese Academy of Sciences, Shanghai, 200030, China}
\address[2]{School of Astronomy and Space Science, University of Chinese Academy of Sciences,
Beijing, 100049, China}
\address[3]{School of Fundamental Physics and Mathematical Sciences, Hangzhou Institute for Advanced Study, UCAS, Hangzhou 310024, China} 
\address[4]{Key Laboratory of Radio Astronomy and Technology, Chinese Academy of Sciences, A20 Datun Road, Chaoyang District, Beijing 100101, China} 
\address[5]{Department of Basic Education, Sanda University, Shanghai 201209, China} 

\abstract{
The origin of supermassive black holes (SMBHs) is a pivotal problem in modern cosmology. This work explores the potential of the Taiji–TianQin space-borne gravitational-wave (GW) detector network to identify the formation channels of massive black hole binaries (MBHBs) at high redshifts ($z \gtrsim 10$). The network substantially improves detection capability, boosting the signal-to-noise ratio by a factor of 2.2–3.0 (1.06–1.14) relative to TianQin (Taiji) alone. It increases the detection rate of MBHBs formed from light seeds (LS) by more than 2.2 times and achieves over 96\% detection efficiency for those originating from heavy seeds (HS). Furthermore, the network enables component mass estimation with relative uncertainties as low as $\sim 10^{-4}$ at the $2\sigma$ level. These improvements facilitate the assembly of a well-constrained population sample, allowing robust measurement of the fractional contributions from different formation pathways. The network achieves high precision in distinguishing between LS and HS origins (7.4\% relative uncertainty at $2\sigma$) and offers moderate discrimination between delay and no-delay channels in HS-origin binaries (24\%). However, classification remains challenging for delay versus no-delay scenarios in LS-origin systems (58\%) due to significant population overlap. In conclusion, the Taiji–TianQin network will serve as a powerful tool for unveiling the origins of SMBHs through GW population studies.
}



\maketitle


\begin{multicols}{2}
\section{Introduction}\label{sec:1}
Recent observations of quasars already at redshifts $z \gtrsim 7$ indicate that supermassive black holes (SMBHs) with masses $ M_{\rm BH} \gtrsim 10^9 ~\msun$  formed very early in the history of the Universe, accumulating their mass within a few hundred million years (Myr)  after the Big Bang~\cite{Inayoshi_2020,Greene_2020,fan2022quasarsintergalacticmediumcosmic,huang2024supermassiveprimordialblackholes,kovacs2024candidatesupermassiveblackhole,maiolino2024smallvigorousblackhole}. 
How did the first SMBHs grow so massive so quickly~\cite{ellis2024originjwstsmbhs,Volonteri_2010,2019_reveal_SMBH_origin}?  It is  widely
\Authorfootnote 
\noindent 
 accepted that SMBHs were not born with such large masses; instead, they must have grown from initial black hole (BH) seeds over cosmic time~\cite{Volonteri:2021sfo}.
Two primary seeding scenarios are commonly considered. The first is the “light seed” (LS) scenario, in which the seeds originate from the remnants of the first Population III (PopIII) stars at redshifts $z \approx 15–20$, with typical masses of $\sim 10^{2}~\msun$ \cite{Klein2016:catalogue,regan2024massiveblackholeseeds}.
The second is the “heavy seed” (HS) scenario, which posits that seeds form via the direct collapse of protogalactic disks at high redshifts ($z \approx 15–20$), resulting in much heavier seeds with masses of $\sim 10^{5}~\msun$~\cite{Klein2016:catalogue,regan2024massiveblackholeseeds,Wise_2019}.

In addition to the seed types, the treatment of the time delay between BH and galaxy mergers—whether it is included (delay, d) or not (no-delay, nod)—also plays a crucial role~\cite{Klein2016:catalogue}.
Following the merger of two galaxies with central BHs, dynamical friction drives the BHs toward the center of the newly formed galaxy, eventually resulting in a bound binary.
Subsequently, during the orbital decay phase, the efficiency of the hardening processes—such as three-body stellar interactions~\cite{Quinlan_1996,10.1093/mnrasl/slv131}, gas-driven migration~\cite{Duffell_2020}, and interactions with other BHs~\cite{Barausse_2020}—remains highly uncertain.
This uncertainty directly impacts the ability of massive black hole binaries (MBHBs) to efficiently transition from parsec(pc)-scale to sub-parsec separations—a challenge known as the ``last parsec problem"~\cite{Milosavljevic2003}.
The physics governing both BH seeding and the time delays between BH and galaxy mergers at high redshift significantly influence the predicted merger rates of MBHBs, as well as the distributions of key parameters such as chirp mass, effective spin, and redshift~\cite{2023_PRD_PTA_LISA_BHseed_Barausse,Barausse_2020,prd_2021_LISA_SMBH}.

At such redshifts or higher ($z\gtrsim10$), electromagnetic observations become increasingly challenging \cite{Volonteri:2021sfo}. Gravitational wave (GW) observatories are crucial in this context, as GW signals decay slowly with redshift and interact weakly with matter. The next generation of ground-based GW observatories, such as the Einstein Telescope (ET) \cite{Abac_2025_ET}, along with space-borne GW detectors like the Laser Interferometer Space Antenna (LISA) \cite{LISA2017,LISA2019PSD}, Taiji \cite{Taiji2017,Taiji2020ruan,du2025Taijichallenge,Ren_2023_TaijiChallenge,zhongxingyu}, and TianQin \cite{Luo_2016_TianQin,tainqin2019,TianQin2021}, will open unprecedented windows into the birth of the first SMBHs~\cite{colpi2019astro2020sciencewhitepaper,Sesana_2011,prd_2021_LISA_SMBH,Hartwig_2018}. Together, they will be uniquely powerful probes of the origin, growth, and formation pathway of SMBHs in the first billion years of the universe.

In this work, we investigate the potential of the joint Chinese Taiji-TianQin network to distinguish between different seeding scenarios.
First, we analyze detection rates of MBHB mergers across different seed models (LS\_d, LS\_nod, HS\_d, HS\_nod) under three observational scenarios: Taiji alone, TianQin alone, and their combined network. 
Our results show that the combined Taiji-TianQin network significantly enhances the SNR for merger signals from MBHBs at high redshifts ($z \gtrsim 10$), achieving an improvement by a factor of 2.2–3.0 (1.06–1.14) relative to TianQin (Taiji) alone.
This improvement enables the detection of faint signals that are undetectable by individual instruments, significantly increasing the number of observable events. In particular, the network greatly enhances the detection of MBHB mergers: for LS models, the detection rate increases by a factor of $2.2$ to $3$  compared to Taiji alone, while TianQin alone cannot detect such events. For HS models, the network achieves a high detection efficiency, capturing more than 96\% of merger events.

Second, we evaluate the effectiveness of the Taiji-TianQin network in extracting physical information from MBHB signals. We assess its parameter estimation capabilities through Bayesian inference using the \texttt{dynesty}~\cite{2020MNRAS4933132S,sergey8408702} nested sampling package. Our analysis focuses on seven key parameters: the component masses ($m_1$, $m_2$), dimensionless spins ($a_1$, $a_2$), spin-orbit tilt angles ($\beta$, $\gamma$), and luminosity distance ($D_L$). These parameters collectively determine the system's logarithmic chirp mass ($\log_{10}\mathcal{M}_c$), inverse hyperbolic tangent of the effective spin ($\mathrm{arcth}(\chi_{\mathrm{eff}})$), and redshift ($z$). The network achieves the following typical relative uncertainties (at the $2\sigma$ level): $m_1$: $10^{-4}$, $m_2$: $10^{-4}$, $a_1$: $10^{-4}$, $a_2$: $10^{-3}$, $\beta$: $10^{-2}$, $\gamma$: $10^{-2}$, and $D_L$: $10^{-4}$.

Finally, we examine whether the Taiji-TianQin network can statistically infer the mixing fractions among different formation channels of BH seeds.
Our analysis follows a sequential two-stage procedure to estimate the formation fractions $f_1$ (LS fraction), $f_2$ (delay fraction in the LS case), and $f_3$ (delay fraction in the HS case).  In the first stage, we constrain the channel fractions using the $\rm{arcth}(\chi_{\rm eff})$ distributions with uniform priors $\mathcal{U}[0,1]$, obtaining posterior distributions well-approximated by Gaussian functions. These posteriors are then used as priors in the second stage, where we incorporate information from $\log_{10}\mathcal{M}_c$, producing refined posterior distributions with significantly reduced uncertainties.
This approach yields three key parameters with distinct precisions: $f_1$ (7.4\% relative uncertainty at 2$\sigma$), $f_2$ (58\%), and $f_3$ (24\%).
The results show that the network achieves high precision in distinguishing between LS and HS origins, offers moderate discrimination for delay versus no-delay channels in HS-origin binaries, and exhibits limited performance in classifying delay scenarios for LS-origin systems—primarily due to significant population overlap between the LS\_d and LS\_nod scenarios.

In more detail, the paper is structured as follows. Sec.~\ref{sec:2} describes the astrophysical models used to generate our MBHB catalog. In Sec.~\ref{sec:3}, we introduce the method for GW data simulation, detection, and parameter estimation with the Taiji-TianQin network. Sec.~\ref{sec:4} details the hierarchical inference methodology for determining formation channel fractions. The conclusion is given in Sec.~\ref{sec:5}.

\section{Simulations of MBHB catalogues}\label{sec:2}
In this section, we present a comprehensive overview of the simulated MBHB catalogs used in our study. These catalogs incorporate two key physical mechanisms that significantly influence merger event rates: (i) black hole seeds and (ii) time delays between galaxy mergers and subsequent black hole mergers. Following the approaches of \cite{Barausse2012:catalogue} (B12), \cite{Klein2016:catalogue} (K16), and \cite{Barausse_2020} (B20), we examine both light-seed and heavy-seed scenarios, as well as delay and no-delay cases for MBHB formation models\footnote{The catalogs are available at \href{https://people.sissa.it/~barausse/catalogs/}{https://people.sissa.it/~barausse/catalogs/}.}. We begin with an introduction to the MBHB seeding models employed in our analysis.

\subsection{Black hole seeds}

A critical component in predicting MBHB merger rates is the initial mass function of high-redshift black hole seeds. Among the many proposed formation mechanisms~\cite{Latif_2016_BHseed,2023_PRD_PTA_LISA_BHseed_Barausse}, we focus on two representative scenarios:

(1) \textbf{Light seed scenario:} Seeds form from the remnants of PopIII stars in low-metallicity environments at $z\approx 15-20$. The initial stellar masses follow a log-normal distribution centered at $300~\msun$ with a standard deviation of 0.2 dex, incorporating a pair-instability gap between $140–260 \msun$ \cite{Heger_2002_popIII,Inayoshi_2020,2023_PRD_PTA_LISA_BHseed_Barausse}. The resulting black hole seed retain approximately 2/3 the mass of the initial star~\cite{Klein2016:catalogue}.

(2) \textbf{Heavy seed scenario:} Seeds originate from the collapse of protogalactic disks induced by bar instability, typically yielding seeds with masses $M_{\rm seed} \sim 10^5~\msun$ predominantly at $z\approx 15-20$.
This formation mechanism is governed by the critical Toomre parameter $Q_c$~\cite{MNRAS_2008_Qc}, which takes physically plausible values in the range $Q_c \approx 1.5-3$. A threshold of $Q_c > 2$ is necessary to ensure that a sufficient fraction of massive galaxies host a massive black hole at $z=0$. The value of $Q_c$ directly determines the seed occupation fraction at high redshifts, with lower values resulting in fewer seeds and higher values producing more seeds~\cite{Klein2016:catalogue,2023_PRD_PTA_LISA_BHseed_Barausse}.

\subsection{Delays}
\begin{figure*}[!ht]
    \centering
    \includegraphics[width=0.9\textwidth]{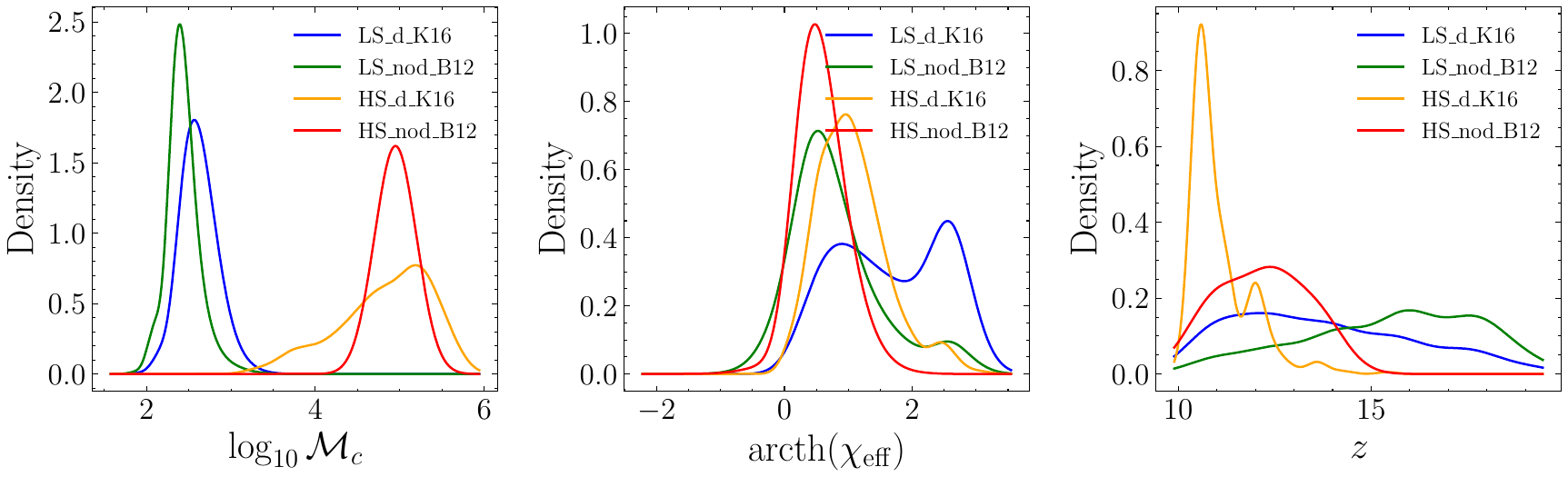}
    \caption{
    Distributions for high-redshift systems ($z \geq 10$) under different formation channels. The parameters shown are the logarithmic chirp mass ($\log_{10}\mathcal{M}_c$), the inverse hyperbolic tangent of the effective spin ($\mathrm{arcth}(\chi_{\mathrm{eff}})$), and the redshift ($z$).
    The blue, green, orange, and red lines represent the LS\_d\_K16, LS\_nod\_B12, HS\_d\_K16, and HS\_nod\_B12 models, respectively.}
    \label{fig:KDE_ALL}
\end{figure*}

The delays between halo/galaxy merger and BH coalescence occur through three sequential phases~\cite{2023_PRD_PTA_LISA_BHseed_Barausse,antoniadis2024seconddatareleaseeuropean}:

\textbf{Halo Merger} (a few Gyr):
When two halos merge according to the dark matter tree, the smaller one survives as a satellite subhalo within the newly formed system and gradually sinks to the center due to dynamical friction ~\cite{Boylan_Kolchin_2008}. This phase typically lasts a few Gyr.

\textbf{Galaxy Merger} (several Gyr; $\sim \rm{kpc}-\rm{pc}$ separation):
After the subhalo reaches the bottom of the host halo, the galaxies do not merge immediately. The satellite galaxy continues to infall toward the center of the host galaxy, driven by dynamical friction and tidal stripping/evaporation ~\cite{Dosopoulou_2017}. This phase can last for several Gyr, especially for galaxies with unequal stellar masses, and is crucial to driving the BH binaries from $\sim$ kpc to $\sim$ pc separation. 

\textbf{BH Binary Evolution} (a few Gyr; sub-pc separation):
Once the galaxies have merged, the two BHs are slowly dragged toward the center of the newly formed galaxy via dynamical friction against the stellar background, eventually forming a bound binary. Subsequently, the binary hardens further through three-body interactions with stars (``stellar hardening"~\cite{Quinlan_1996}), typically reaching the small separations ($\lsim 10^{-2}$ pc) within a few Gyr. Additional mechanisms—such as galaxy rotation~\cite{Holley_Bockelmann_2015,Mirza_2017}, planetary-like migration~\cite{10.1111/j.1365-2966.2009.15179.x}, and triple MBH interactions~\cite{2007_tripleBH}—may also facilitate the binary's evolution into the sub-pc separations ~\cite{Colpi_2014}.
The efficiency of these processes remains significantly uncertain; however, they play a critical role in influencing the capacity of MBHBs to transition effectively from parsec-scale to sub-parsec separations — a challenge known as the ``last parsec problem"~\cite{Milosavljevic2003}.

The total delay between the initial halo merger and the final BH coalescence reflects the cumulative duration of these three evolutionary phases. This hierarchical delay structure shapes the predicted merger rates and the distributions of parameters like chirp mass, effective spin, and redshift. In the absence of such delays, BH binaries merge promptly before they have a chance to grow and substantially spin up~\cite{Barausse_2020}.
The specific delay models considered can be broadly categorized into three classes~\cite{2023_PRD_PTA_LISA_BHseed_Barausse}. All incorporate delays between halo and galaxy mergers, but they differ in the degree of delay between galaxy and black hole mergers: 
(i) \textbf{No delays}: These assume the BH mergers at the same time as their host galaxies. Representative examples include LS\_nod (B12~\cite{Barausse2012:catalogue}), HS\_nod (B12), and Q3\_nod (K16~\cite{Klein2016:catalogue}).
(ii) \textbf{Medium delays}: These incorporate additional processes at sub-pc separation, such as stellar hardening,  MBH triplets, and planetary-like migration. Representative cases are popIII\_d (K16~\cite{Klein2016:catalogue}) and Q3\_d (K16~\cite{Klein2016:catalogue}).
(iii) \textbf{Long delays}: These account for the complete hierarchical delay, including the evolution of black hole pairs at separations of $\sim$ 100 pc. This category includes LS\_SN\_d (B20~\cite{Barausse_2020,antoniadis2024seconddatareleaseeuropean}), LS\_noSN\_d (B20~\cite{Barausse_2020,antoniadis2024seconddatareleaseeuropean}), HS\_SN\_d (B20~\cite{Barausse_2020,antoniadis2024seconddatareleaseeuropean}), and HS\_noSN\_d (B20~\cite{Barausse_2020,antoniadis2024seconddatareleaseeuropean}). 
The labels ``SN" and ``noSN" indicate simulations with and without supernova (SN) feedback, respectively; ``LS" (or ``popIII") and ``HS" (or ``Q3") denote the light-seed and heavy-seed scenarios for the high-redshift initial mass function.

We consider in particular the semianalytic model in B12 (no delay) and K16 (medium delay).
The four specific model combinations used in this work are:
\begin{itemize}
  \item[(1)] \textbf{LS\_d} (popIII\_d in K16~\cite{Klein2016:catalogue}): A light-seed scenario in which the MBH seed originates from the remnants of the first generation (PopIII) stars, with a typical mass of $\sim100~\msun$. This model also incorporates time delays between galaxy and BH mergers.

  \item[(2)] \textbf{HS\_d} (Q3\_d in K16~\cite{Klein2016:catalogue}): A heavy-seed scenario in which the MBH seed forms via the direct collapse of protogalactic disks, with a mass of $\sim10^5~\msun$. This model also includes the time delays between galaxy and BH mergers, with the critical Toomre parameter set to $Q_c = 3$.

  \item[(3)] \textbf{LS\_nod} (B12~\cite{Barausse2012:catalogue}): A light-seed scenario similar to LS\_d, but assume
no delays between galaxy and BH mergers.

  \item[(4)] \textbf{HS\_nod} (B12~\cite{Barausse2012:catalogue}): A heavy-seed scenario similar to HS\_d, but the time delay between galaxy and BH mergers is not included.
\end{itemize}

\section{GW data simulation, detection, and parameter estimation}\label{sec:3}

\subsection{GW data simulation}

\begin{table}[H]
\footnotesize
\begin{threeparttable}\caption{Summary of GW parameters and their meaning included in each catalog.}\label{tab:catalogs}
\doublerulesep 0.1pt \tabcolsep 0.2pt 
\begin{tabular}{cc}
\toprule
  Parameter & Description 
  \\\hline
  $z$   & redshift     \\
  $m_1$&  the component black hole mass     \\
  $m_2$&  the component black hole mass  \\
  $a_1$ & dimensionless spin magnitude  \\
  $a_2$ & dimensionless spin magnitude \\
  $\alpha$ &the angle between the spin vectors $\hat{a}_1$ and $\hat{a}_2$ \\
  $\beta$ & the angle between $\hat{a}_1$ and the orbital angular momentum $\hat{L}$ \\
  $\gamma$ &  the angle between $\hat{a}_2$ and $\hat{L}$ \\
  $\psi$ & the angle between the projections of $a_1$ and $a_2$ onto the orbital plane \\
\bottomrule
\end{tabular}
\end{threeparttable}
\end{table}

We simulate MBHB merger signals from the four catalogs described in Sec.~\ref{sec:2} using the phenomenological waveform model \texttt{IMRPhenomPv2} \cite{Hannam_2014,Khan_2019,jiangye} implemented in \texttt{PyCBC}~\cite{alex_nitz_2024_pycbc}.
As shown in Tab.~\ref{tab:catalogs}, 
each catalog includes the following GW parameters: the redshift $z$ of the BH merger; the component black hole masses $m_1$ and $m_2$ (in solar masses); the dimensionless spin magnitudes $a_1$ and $a_2$; the angle $\alpha$ between the spin vectors $\hat{a}_1$ and $\hat{a}_2$, defined via $\hat{a}_1 \cdot \hat{a}_2 = \cos \alpha$; the angle $\beta$ between $\hat{a}_1$ and the orbital angular momentum $\hat{L}$, defined by $\hat{a}_1 \cdot \hat{L} = \cos \beta$; and the angle $\gamma$ between $\hat{a}_2$ and $\hat{L}$, with $\hat{a}_2 \cdot \hat{L} = \cos \gamma$; and the angle $\psi$ between the projections of $\hat{a}_1$ and $\hat{a}_2$ onto the binary orbital plane.

In this work, we focus exclusively on MBHBs at high redshift ($z \geq 10$). The chirp mass $\mathcal{M}_c$, effective spin $ \chi_{\rm eff}$, and its inverse hyperbolic tangent transformation $\mathrm{arcth}(\chi_{\mathrm{eff}})$ are defined as:
\begin{equation}
    \mathcal{M}_c=(m_1m_2)^{3/5}/(m_1+m_2)^{1/5},
\end{equation}
\begin{equation}
    \chi_{\rm eff}=\frac{m_1 a_1 \cos(\beta)+m_2 a_2 \cos(\gamma)}{m_1+m_2}, 
\end{equation}
\begin{equation}
    \mathrm{arcth}(\chi_{\mathrm{eff}}) = \frac{1}{2} \ln\left(\frac{1 + \chi_{\mathrm{eff}}}{1 - \chi_{\mathrm{eff}}}\right), \quad |\chi_{\mathrm{eff}}| < 1
\end{equation}
where $\mathcal{M}_c$ traces the black hole seed mass (distinguishing light from heavy seeds), while $ \chi_{\rm eff}$ quantifies the effective spin-orbit alignment and encodes the formation history of the binary system \cite{PRL_2025_BHspin}.
The inverse hyperbolic tangent of the effective spin, $\mathrm{arcth}(\chi_{\mathrm{eff}})$, provides enhanced sensitivity to spin orientation effects: values clustered near zero ($\chi_{\mathrm{eff}} \approx 0$) indicate randomly oriented spins, while significant deviations ($\chi_{\mathrm{eff}}\gtrsim 0.76$ corresponding to $\mathrm{arcth}(\chi_{\mathrm{eff}})\gtrsim1$) reveal strong alignment mechanisms.

Fig.~\ref{fig:KDE_ALL} presents the population distributions of  $\log_{10}\mathcal{M}_c$, $\rm{arcth}(\chi_{\rm eff})$, and $z$, derived through kernel density estimation (KDE) \cite{Scott2012} using \texttt{SciPy} \cite{2020SciPy-NMeth}. 
The $\log_{10}\mathcal{M}_c$ distribution (left panel) reveals a distinct separation between LS and HS populations. The HS distribution exhibits greater variance, suggesting diverse formation pathways potentially involving multiple accretion episodes or hierarchical mergers.

The $\mathrm{arcth}(\chi_{\mathrm{eff}})$ distributions (middle panel) exhibit different characteristics between delay and no-delay cases. Here, $\chi_{\mathrm{eff}} \in (-1,1)$ represents the effective spin parameter quantifying spin-orbit alignment, while its $\mathrm{arcth}$ transformation serves to enhance sensitivity near extreme values. 
The no-delay models (LS\_nod, HS\_nod) peak near zero, consistent with weakly aligned or isotropic spin orientations. In contrast, models incorporating delay mechanisms (LS\_d, HS\_d) exhibit pronounced peaks at $\mathrm{arcth}(\chi_{\mathrm{eff}}) \gtrsim 1$, indicating strongly aligned spins. This distinction arises because, without delays, BHs merge promptly before significant gas-driven spin-up can occur, whereas delay mechanisms allow time for accretion-driven alignment and spin amplification~\cite{Barausse_2020}.

The redshift distributions (right panel) show that delayed-formation models peak systematically at lower redshifts than their non-delayed counterparts, owing to the additional time introduced by the delay.
Note that Fig.~\ref{fig:KDE_ALL} reflects the intrinsic population distributions, without accounting for detection thresholds. In subsequent sections, we will present the detectable population (signal-to-noise ratio, $\mathrm{SNR} \geq 8$), whose distributions are anticipated to differ significantly.

\subsection{Detection}\label{sec:Detectable rates}
\begin{figure}[H]
    \centering
    \includegraphics[width=0.45\textwidth]{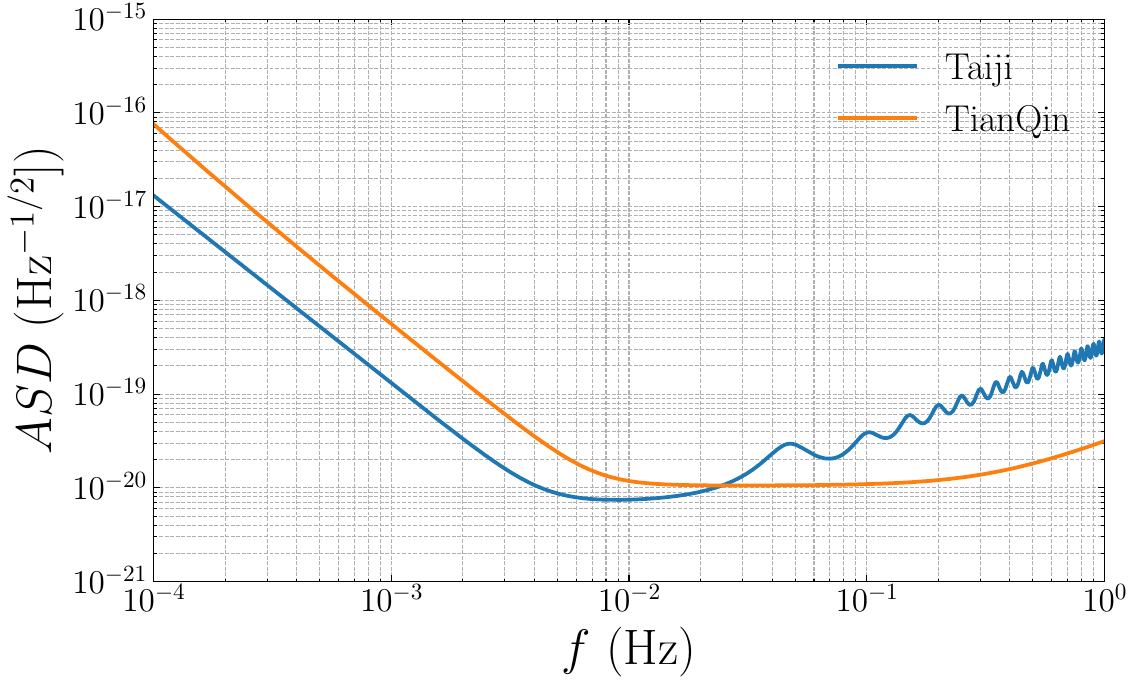}
    \caption{The sensitivity curves of Taiji~\cite{Taiji2020ruan} and TianQin~\cite{tainqin2019}.}
    \label{fig:psd}
\end{figure}

In general, the observed signal $d(t)$ can be expressed as the sum of the GW signal $h(t)$ and the detector noise $n(t)$. 
For a single detector, the SNR of a given source is defined as $\sqrt{\langle d|h\rangle}$, where the noise-weighted inner product is given by
\begin{equation}\label{eq:inner_product}
\langle a|b\rangle \equiv 2 \int_0^{\infty} df
\frac{a^*(f)b(f) + a(f)b^*(f)}{S_n(f)},
\end{equation}
with $S_n(f)$ denoting the power spectral density (PSD) of the detector noise $n(t)$. We generate the detector noise using the $S_n(f)$, adopting the Taiji noise model from Ref.~\cite{Taiji2020ruan} and the TianQin noise model from Ref.~\cite{tainqin2019}.
The corresponding amplitude spectral densities (ASDs), derived from $\sqrt{S_n(f)}$, are shown in Fig.~\ref{fig:psd}.

For a network of $N$ detectors, current data-analysis pipelines typically adopt coincidence analysis rather than a coherent approach~\cite{Sathyaprakash_2009}.
In the coherent approach, data from all detectors are coherently combined and filtered using a consistent signal model, whereas in the coincidence approach, each detector’s data are analyzed independently and cross-compared to identify coincident events.
This preference arises primarily from the nature of detector noise and computational feasibility~\cite{Sathyaprakash_2009}.
Firstly, the noises of different detectors are dominated by independent instrumental sources
and are neither Gaussian nor stationary. Under these conditions, the assumption of statistical independence between detectors is well justified, and
 coincidence analysis can potentially reduce the background. 
Secondly, coherent analysis is computationally far more expensive than coincidence analysis, making it presently impractical to employ~\cite{Sathyaprakash_2009}.
Under the coincidence framework, the network SNR for $N$ detectors is given by~\cite{Jin_2023_Hubble_3network}:
\begin{equation}
    \rho_{ N} = \sqrt{\sum_{i=1}^{N} \rho_i^2},
\label{eq:rho}
\end{equation}
which, for the Taiji–TianQin network, reduces to
\begin{equation}
\rho_{\rm joint} = \sqrt{\rho_{\rm Taiji}^2 + \rho_{\rm TianQin}^2}.
\label{eq:1}
\end{equation}

\begin{figure}[H]
    \centering
    \includegraphics[width=0.4\textwidth]{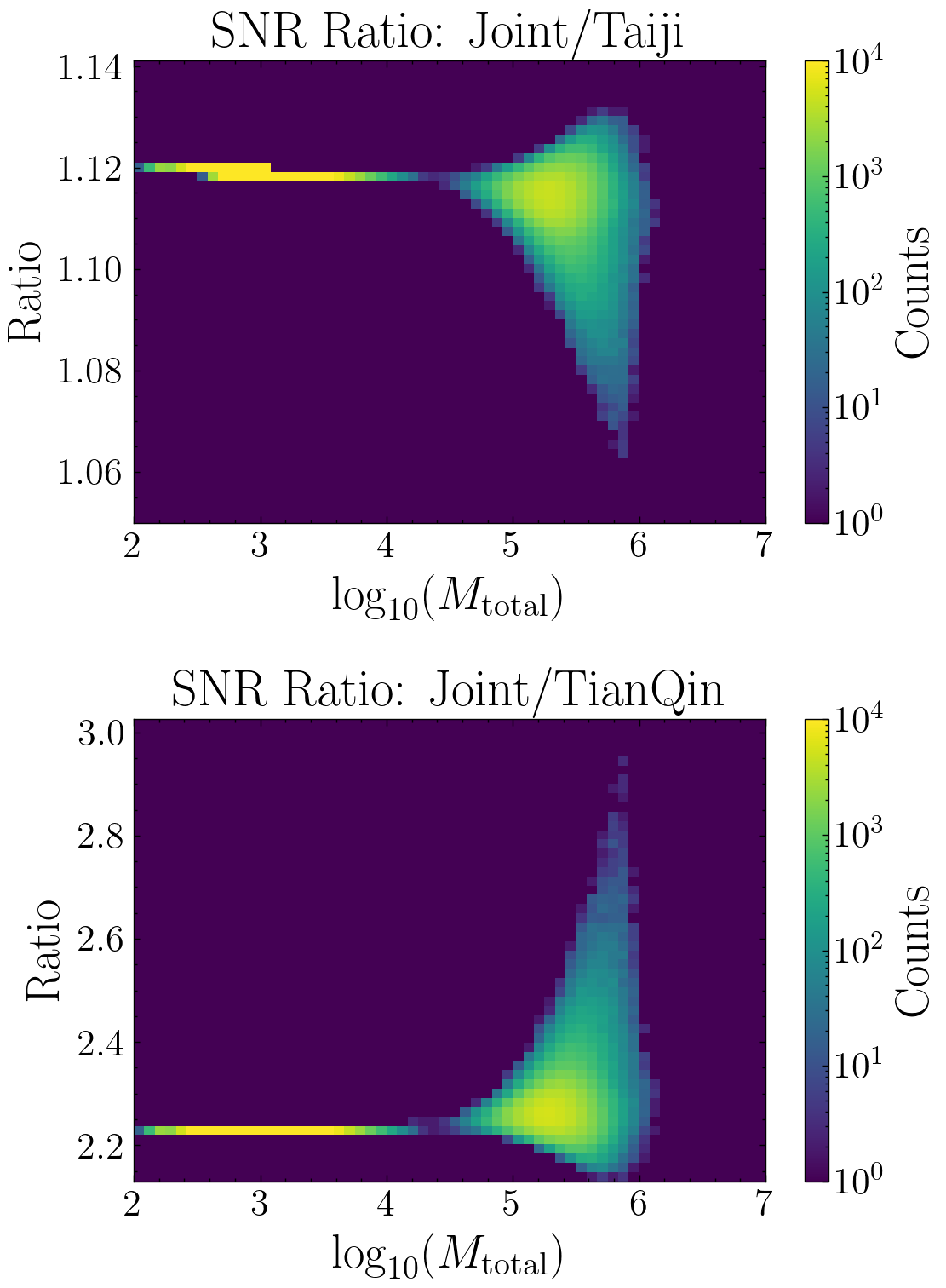}
    \caption{SNR enhancement from Taiji-TianQin network. Top: Ratio $\rho_{\rm joint}/\rho_{\rm Taiji}$ versus total mass. Bottom: Corresponding ratio $\rho_{\rm joint}/\rho_{\rm TianQin}$. The colour bar represents the number of simulated binaries. Here, all four seeding models (LS\_d, LS\_nod, HS\_d, and HS\_nod) are included.}
    \label{fig:SNR}
\end{figure}

Fig.~\ref{fig:SNR} compares the SNR achieved by the joint Taiji–TianQin network with that of each detector operating individually. The top and bottom panels show the ratios $\rho_{\mathrm{joint}}/\rho_{\mathrm{Taiji}}$ and $\rho_{\mathrm{joint}}/\rho_{\mathrm{TianQin}}$ as functions of total mass, respectively. 
The network significantly enhances the SNR for MBHB merger signals, achieving an improvement by a factor of 2.2–3.0 (1.06–1.14) relative to TianQin (Taiji) alone.
This difference arises from Taiji’s complementary sensitivity, which compensates for TianQin’s higher noise floor in specific frequency bands.

For detectability, we adopt an SNR threshold of 8. 
Tab.~\ref{tab:detector_rates} represents the number of detectable GW events ($N_{\rm det}$) for each model (LS\_d, LS\_nod, HS\_d, HS\_nod) under three detection scenarios: Taiji, TianQin, and their combined network.
Our results demonstrate that the Taiji-TianQin network significantly enhances detection capabilities. For the LS models (LS\_d and LS\_nod), the network increases the detection rate by a factor of 2.2 to 3 compared to Taiji alone. TianQin alone, conversely, cannot detect such events. For the HS models (HS\_d and HS\_nod), the network achieves high detection efficiency, capturing over 96\% of all merger events. This enhancement enables the detection of previously inaccessible faint signals, thereby expanding the observable population of MBHBs. The substantially larger event sample provided by the network is crucial for performing population-level studies, which will allow us to reliably infer the origin of SMBHs.

\subsection{Parameter estimation }\label{sec:Parameter estimation}
Given a simulated observed signal $d(t) = h(t) + n(t)$, where the noise $n(t)$ is generated according to the PSD described in Section~\ref{sec:Detectable rates}, we employ the \texttt{IMRPhenomPv2} waveform model for parameter recovery~\cite{yunqianyun}. In this analysis, we assume the source has already been identified in the data stream and focus exclusively on parameter estimation. The likelihood is given by
\begin{equation}
\mathcal{L}(\theta) \propto \exp[-\frac{1}{2} \langle d-h(\theta)|d-h(\theta) \rangle]\, ,
\end{equation}
where the noise-weighted inner product is defined as Eq.~\ref{eq:inner_product}.
The injected parameters are $z_{\rm inj}=11.0979$ (corresponding to ${D_L}_{\rm inj}=119426\  \rm{Mpc}$), with component masses ${m_1}_{\rm inj}=461251\msun$ and ${m_2}_{\rm inj}=89022.5\msun$, and spins ${a_1}_{\rm inj}=0.851959$ and  ${a_2}_{\rm inj}=0.844949$. Additionally, the angles $\beta_{\rm inj}=0.069212$, $\gamma_{\rm inj}=0.162004$, and $\psi_{\rm inj}=3.84752$ are included.

We perform Bayesian inference using the nested sampling package \texttt{dynesty}~\cite{2020MNRAS4933132S,sergey8408702} to obtain posterior probability distributions.
Our analysis focuses on a set of parameters $\{m_1, m_2, a_1, a_2, \beta, \gamma, D_L\}$ for the binary system, with direct relevance to its chirp mass ($\mathcal{M}_c$), effective spin ($\chi_{\rm eff}$), and redshift ($z$).
 For computational efficiency while maintaining physical relevance, we adopt narrow Gaussian priors on all parameters (see Tab.~\ref{tab:prior_posterior_PE}). This approach allows us to focus on the influence of key parameters in distinguishing formation channels (see Sec.~\ref{sec:4}), while a more comprehensive parameter estimation study across the full parameter space is reserved for future work.
 
 \begin{table}[H]
\footnotesize
\begin{threeparttable}\caption{Number of total GW events ($N_{\rm tot}$) and detectable events ($N_{\rm det}$) for various models and detection scenarios at high redshift ($z \gtrsim 10$). Percentages of detectable events are shown in parentheses.}\label{tab:detector_rates}
\doublerulesep 0.1pt \tabcolsep 2.6pt 
\begin{tabular}{lcccc}
\toprule
  Type & $N_{\rm tot}$ &  $N_{\rm det(TianQin)}$ &$N_{\rm det(Taiji)}$  & $N_{\rm det(Taiji+TianQin)}$  \\\hline
  LS\_d    & 472578        & 0 (0\%) & 31 (0.0066\%)     & 68  (0.0144\%)  \\
  LS\_nod & 1659482       & 0  (0\%)   & 22 (0.0013\%)    &  66  (0.0040\%) \\
  HS\_d    & 319            & 279 (87.5\%)  & 302  (94.7\%)  &  307  (96.2\%)\\
  HS\_nod & 207266     & 207257 (99.996\%) & 207266 (100\%) & 207266 (100\%) \\ 
\bottomrule
\end{tabular}
\end{threeparttable}

\end{table}
\begin{table}[H]
\footnotesize
\begin{threeparttable}\caption{Prior distributions of parameters $\theta_j$ and their posterior $\mathcal{P}$ with $2\sigma$ (95\%) credible intervals. The notation $\mathcal{G}(\mu,2\sigma)$ denotes a Gaussian distribution with mean $\mu$ and standard deviation $\sigma$.}\label{tab:prior_posterior_PE}
\doublerulesep 0.1pt \tabcolsep 2pt 
\begin{tabular}{llllll}
\toprule
$\theta_j$   & inject &   Prior         & $\mathcal{P_{\rm TianQin}}$  &$\mathcal{P}_{\rm Taiji}$  & $\mathcal{P}_{\rm Join}$  \\
\midrule
$m_1$ & $461251$ & $\mathcal{G}({m_1}_{\rm inj},100)$ & $461227^{+79}_{-72}$ &  $461227^{+76}_{-78}$ &  $461260^{+77}_{-79}$\\
$m_2$ & $89022.5$ &  $\mathcal{G}({m_2}_{\rm inj},100)$ & $88990^{+38}_{-37}$ & $89030^{+16}_{-15}$ & $89030^{+16}_{-16}$  \\
$a_1$ & $0.851959$ &  $\mathcal{G}({a_1}_{\rm inj},0.004)$  & $0.8519^{+0.0006}_{-0.0006}$ &$0.8521^{+0.0005}_{-0.0005}$  & $0.8521^{+0.0004}_{-0.0005}$ \\
$a_2$ &  $0.844949$ &   $\mathcal{G}({a_2}_{\rm inj},0.004)$  & $0.8447^{+0.0033}_{-0.0033}$ & $0.8448^{+0.0034}_{-0.0032}$  & $0.8448^{+0.0035}_{-0.0032}$  \\
$\beta$ &$0.069212 $&  $\mathcal{G}(\beta_{\rm inj},0.004)$   &$0.0710^{+0.0020}_{-0.0021}$  &$0.0700^{+0.0022}_{-0.0022}$  & $0.0706^{+0.0018}_{-0.0019}$  \\    
$\gamma$ &$0.162004$ &  $\mathcal{G}(\gamma_{\rm inj},0.004)$   &$0.1620^{+0.0034}_{-0.0034}$  & $0.1621^{+0.0032}_{-0.0034}$  & $0.1621^{+0.0033}_{-0.0034}$  \\
$D_L$ & $119426$ &  $\mathcal{G}({D_L}_{\rm inj},100)$   &$119414^{+85}_{-79}$ & $119403^{+81}_{-73}$ &  $119396^{+80}_{-69}$ \\ 
\bottomrule
\end{tabular}
\end{threeparttable}
\end{table}

 \begin{figure}[H]
    \centering
    \includegraphics[width=0.45\textwidth]{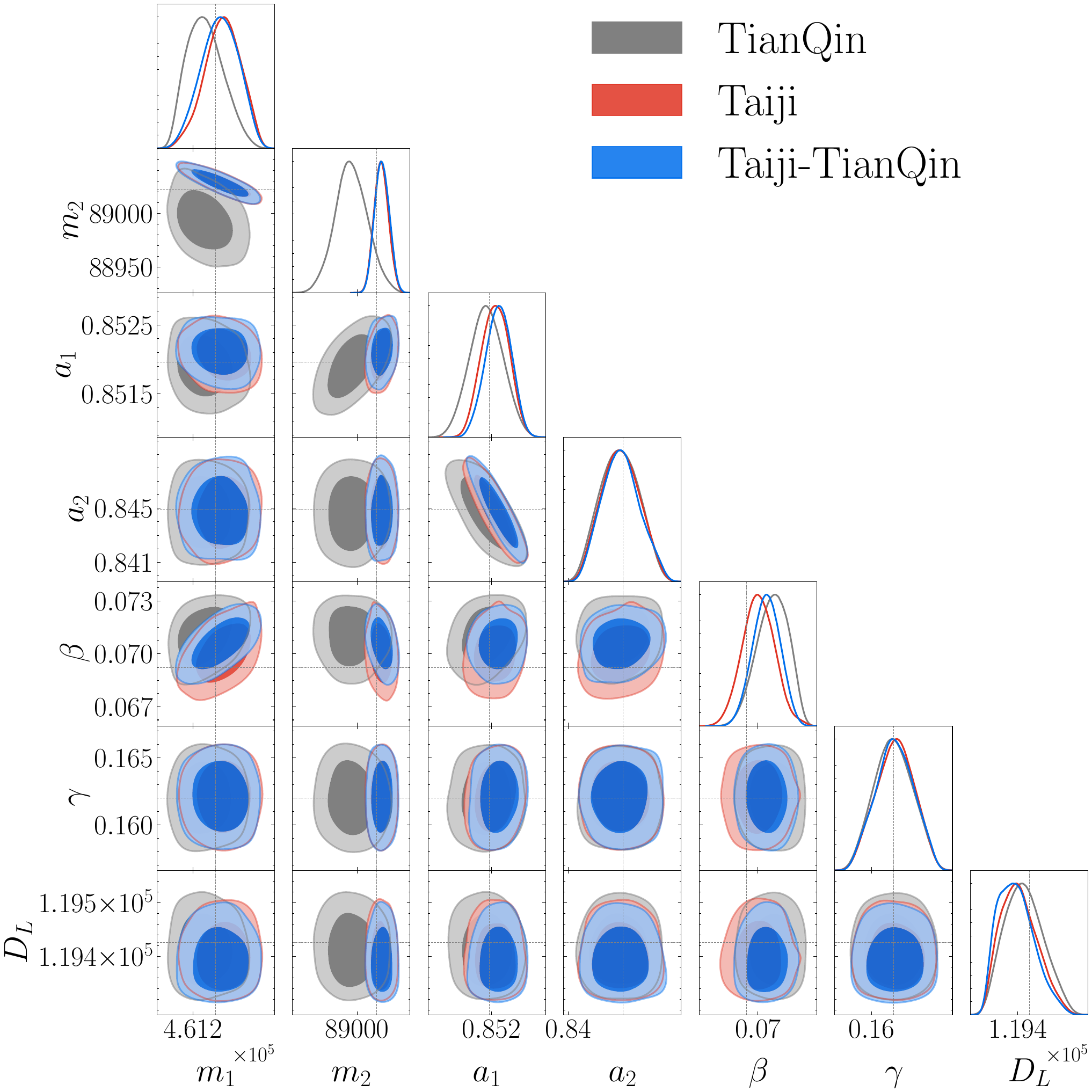}
    \caption{
     Posterior distributions of the source parameters for different detection strategies. The grey, red, and blue curves correspond to the TianQin, Taiji, and the joint Taiji-TianQin network detection scenarios, respectively. The injected values are marked by the vertical black lines.
    }
    \label{fig:PEsmall}
\end{figure} 

\begin{figure*}[htb]
    \centering
    \includegraphics[width=0.9\textwidth]{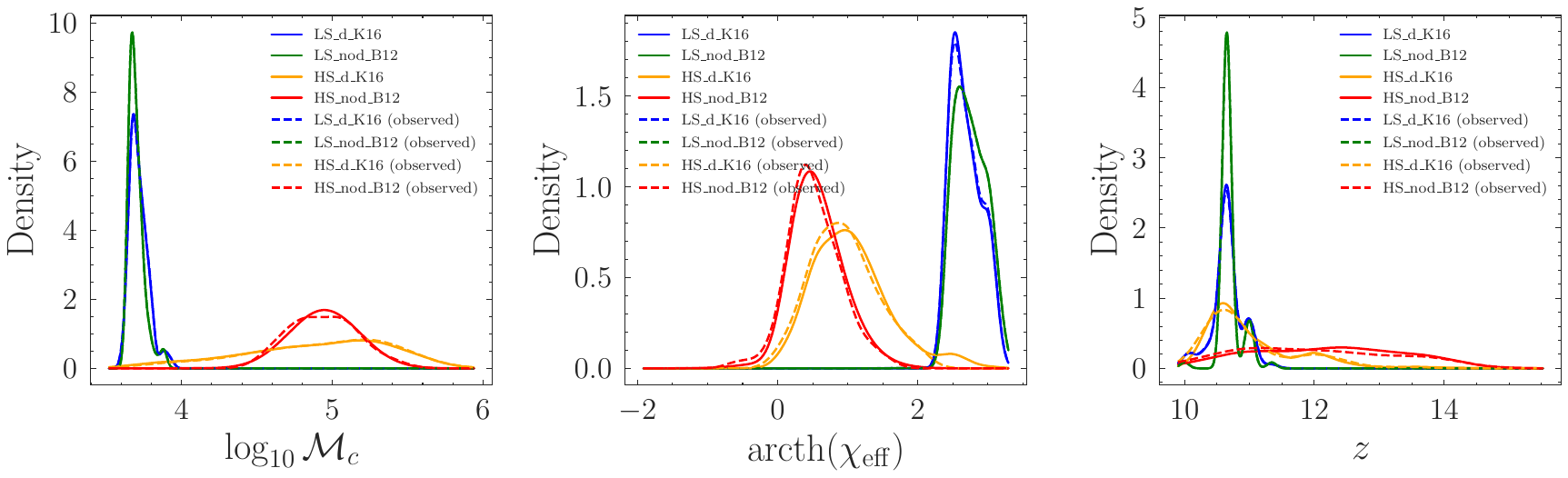}
    \caption{
    Probability density distributions comparing the full detectable set (207,707 events; solid lines) with the observed set (252 events; dashed lines). The three parameters shown are: logarithmic chirp mass ($\log_{10}\mathcal{M}_c$), inverse hyperbolic tangent of the effective spin ($\mathrm{arcth}(\chi_{\mathrm{eff}})$), and redshift ($z$). The blue, green, orange, and red lines represent the LS\_d\_K16, LS\_nod\_B12, HS\_d\_K16, and HS\_nod\_B12 models, respectively.
    }
    \label{fig:KDE_snr8}
\end{figure*}

\begin{figure*}[htb]
    \centering
    \includegraphics[width=0.9\textwidth]{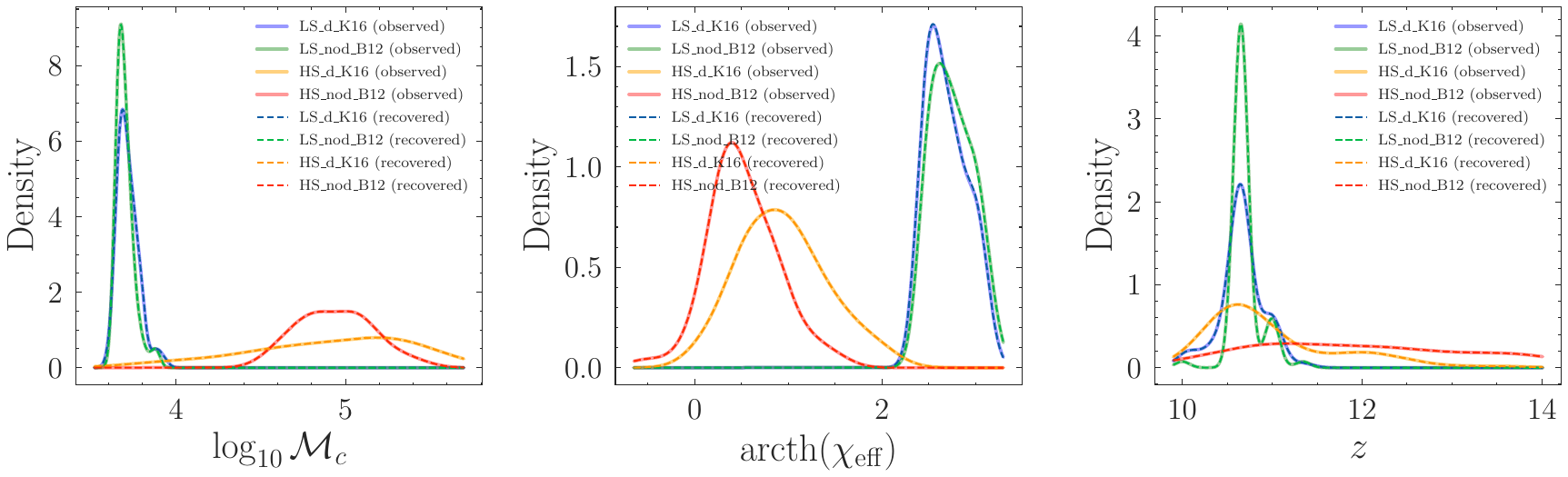}
    \caption{Similar to Fig.~\ref{fig:KDE_snr8} but comparing the observed set (252 events; solid lines) with the recovered results (dashed lines).}
    \label{fig:252_VS_recovered}
\end{figure*}

Fig.~\ref{fig:PEsmall} shows the posterior distributions obtained with TianQin (grey), Taiji (red), and the Taiji–TianQin network (blue), respectively.
Both Taiji and the network yield distributions whose peaks are closer to the true values than those from TianQin. Furthermore, the Taiji-TianQin network produces noticeably narrower posteriors, as seen in the tighter contours, reflecting higher precision in parameter estimation.
Tab.~\ref{tab:prior_posterior_PE} lists the prior distributions used for parameters $\theta_j$ and their corresponding posterior $2\sigma$ (95\%) credible intervals.
The Taiji-TianQin network achieves the following typical relative precisions (at the $2\sigma$ level): $m_1$: $10^{-4}$, $m_2$: $10^{-4}$, $a_1$: $10^{-4}$, $a_2$: $10^{-3}$, $\beta$: $10^{-2}$, $\gamma$: $10^{-2}$, and $D_L$: $10^{-4}$.

\section{Formation channels distinction using Taiji-TianQin network}\label{sec:4}

As shown in Tab.~\ref{tab:detector_rates}, the Taiji-TianQin network can detect 207,707 GW events, constituting our complete \textbf{detectable set}.
Each formation channel simulated in Sec.~\ref{sec:2}  considered only a single seeding prescription. However, the actual population of BHs in the universe is unlikely to be described by such ``pure'' models, but rather by a mixture of different formation scenarios.
To account for this, we introduce three mixing fractions:
 $f_1$ between the LS and HS scenarios, $f_2$ between the LS\_d and LS\_nod sub-channels, and $f_3$ between the HS\_d and HS\_nod sub-channels. 
The complete binary BH population distribution is then modeled as:

\begin{align}
    P= & f_1\times[f_2 \times P_{\rm LS\_d}+ (1-f_2) \times P_{\rm LS\_nod}] \nonumber\\
      &+(1-f_1)\times [f_3\times P_{\rm HS\_d}+ (1-f_3) \times P_{\rm HS\_nod}]
      \label{eq:channal_factor}
\end{align}

In this section, we will distinguish the formation channels from the MBHB populations using the Taiji-TianQin network.
To balance computational efficiency with statistical reliability, we randomly subsampled 252 events (0.12\% of the detectable set) to form the \textbf{observed set}. 
The number and fraction of observed events from each formation channel are summarized in Tab.~\ref{tab:inject_events}.

Fig.~\ref{fig:KDE_snr8} validates our sampling method through the KDE of three parameters: $\mathrm{arcth}(\chi_{\mathrm{eff}})$, $\log_{10}\mathcal{M}_c$, and redshift ($z$, derived from $D_L$). It compares the full detectable population (207,707 events; solid lines) with the observed subset (252 events; dashed lines). The close alignment of the KDE curves indicates that the sample retains its key statistical properties across all channel models (LS\_d\_K16, LS\_nod\_B12, HS\_d\_K16, HS\_nod\_B12), demonstrating the robustness of our selection.

Following the parameter recovery method described in Sec.~\ref{sec:Parameter estimation}, we use \texttt{dynesty} to estimate the parameter set $\bm{\theta} = \{m_1, m_2, a_1, a_2, \beta, \gamma, D_L\}$ for each event in the dataset $\mathcal{D} = \{d_i\}_{i=1}^{252}$, and derive the distributions of $\log_{10}\mathcal{M}_c$, $\mathrm{arcth}(\chi_{\mathrm{eff}})$, and redshift $z$.
As shown in Fig.~\ref{fig:252_VS_recovered}, the reconstructed distributions (dashed curves) agree closely with the true injected distributions (solid curves), confirming accurate recovery of the intrinsic population properties. This consistency across different formation channels confirms the reliability of our method in the presence of potential systematic biases.

We now infer the formation channel fractions for the 252-event sample. Our inference framework models the likelihood of the observing dataset $\mathcal{D} = \{d_i\}_{i=1}^{252}$ through a hierarchical mixture of formation channels. 
The likelihood of the observing  dataset $\mathcal{D}$  for the parameter $\theta_j$ is the product of probabilities for all 252 events:
\begin{equation}
    \mathcal{L}(f_1, f_2, f_3 |\mathcal{D},\theta_j) = \prod_{i=1}^{252} P(d_i | f_1, f_2, f_3, \theta_j),
\end{equation}
where the probability for each event  \( P(d_i | f_1, f_2, f_3, \theta_j) \) is defined as a weighted mixture over four formation channels:
\begin{align}
&P(d_i | f_1, f_2, f_3, \theta_j) =   \nonumber \\
&f_1 \times \left[ f_2 \times P_{\text{LS\_d}}(d_i| \theta_j) + (1-f_2) \times P_{\text{LS\_nod}}(d_i| \theta_j) \right] \nonumber\\
&+ (1-f_1) \times \left[ f_3 \times P_{\text{HS\_d}}(d_i| \theta_j) + (1-f_3) \times P_{\text{HS\_d}}(d_i | \theta_j) \right] .
      \label{eq:probability}
\end{align}

For each event $d_i$, the probability of arising from a specific channel depends on: (1) the population-level fractions $f_1$ (LS vs. HS), $f_2$ (LS\_d vs. LS\_nod), and $f_3$ (HS\_d vs. HS\_nod); (2) the channel-specific probability densities $P_{\text{model}}(d_i | \theta_j)$, where $\rm{ model \in \{LS\_d, LS\_nod, HS\_d, HS\_d \} }$, derived from the full detectable population (207,707 events) using kernel density estimation. 

Tab.~\ref{tab:prior_posterior} presents the hierarchical Bayesian framework used to infer formation channel proportions. The analysis undergoes a two-stage sequential updating process, with each stage refining the posterior distributions:

(1)  \textbf{Effective spin analysis}: We first constrain the channel fractions $(f_1, f_2, f_3)$ using the $\rm{arcth}(\chi_{\rm eff})$ distributions. Beginning with uniform priors $\mathcal{U}[0,1]$ for all fractions, we obtain posterior distributions well-approximated by Gaussian functions $\mathcal{G}(\mu,2\sigma)$ with mean $\mu$ and standard deviation $\sigma$.
    
(2)  \textbf{Chirp mass refinement}: The approximate Gaussian posteriors from the first step become the new priors as we incorporate $\log_{10}\mathcal{M}_c$ information. This stage produces updated posteriors with reduced uncertainties, yielding the final estimation of $(f_1, f_2, f_3)$.

\begin{table}[H]
\footnotesize
\begin{threeparttable}\caption{The number and percentage of observed events across different formation channels.}\label{tab:inject_events}
\doublerulesep 0.2pt \tabcolsep 25pt 
\begin{tabular}{lll}
\toprule
  Type & Number & Percentage \\\hline
  LS\_d    & 66    & 26.2\% \\
  LS\_nod & 66    & 26.2\% \\
  HS\_d   & 59    & 23.4\% \\
  HS\_nod & 61    & 24.2\%\\
\bottomrule
\end{tabular}
\end{threeparttable}
\end{table}

\begin{table}[H]
\footnotesize
\begin{threeparttable}\caption{Strategies and results for formation channel fraction inference. True values of each fraction are indicated in bold.
Posterior distributions are summarized using $2\sigma$ (95\%) credible intervals.
$\mathcal{U}[a,b]$ denotes uniform distribution on $[a,b]$; $\mathcal{G}(\mu,2 \sigma)$ denotes Gaussian distribution with mean $\mu$ and standard deviation $\sigma$. }\label{tab:prior_posterior}
\doublerulesep 0.1pt \tabcolsep 2.5pt 
\begin{tabular}{ll|llll}
\toprule
Step & $\theta_j$      &          & $f_1$ (\textbf{0.524}) & $f_2$ (\textbf{0.5})    & $f_3$ (\textbf{0.492})  \\
\midrule
1    & $\rm{arcth}(\chi_{\rm eff})$ & Prior    & $\mathcal{U}[0,1]$  & $\mathcal{U}[0,1]$   & $\mathcal{U}[0,1]$   \\
     &                  & Posterior& $0.517^{+0.068}_{-0.069}$  & $0.54^{+0.43}_{-0.46}$   & $0.39^{+0.21}_{-0.20}$   \\
\hline
2    & $\log_{10}\mathcal{M}_c$ & Prior    & $\mathcal{G}(0.517,0.069)$  & $\mathcal{G}(0.54,0.46)$   & $\mathcal{G}(0.39,0.21)$    \\
     &                        & Posterior& $0.523^{+0.039}_{-0.039}$    & $0.39^{+0.29}_{-0.26}$    & $0.41^{+0.12}_{-0.12}$ \\
\bottomrule
\end{tabular}
\end{threeparttable}
\end{table}

\begin{figure}[H]
    \centering
    \includegraphics[width=0.42\textwidth]{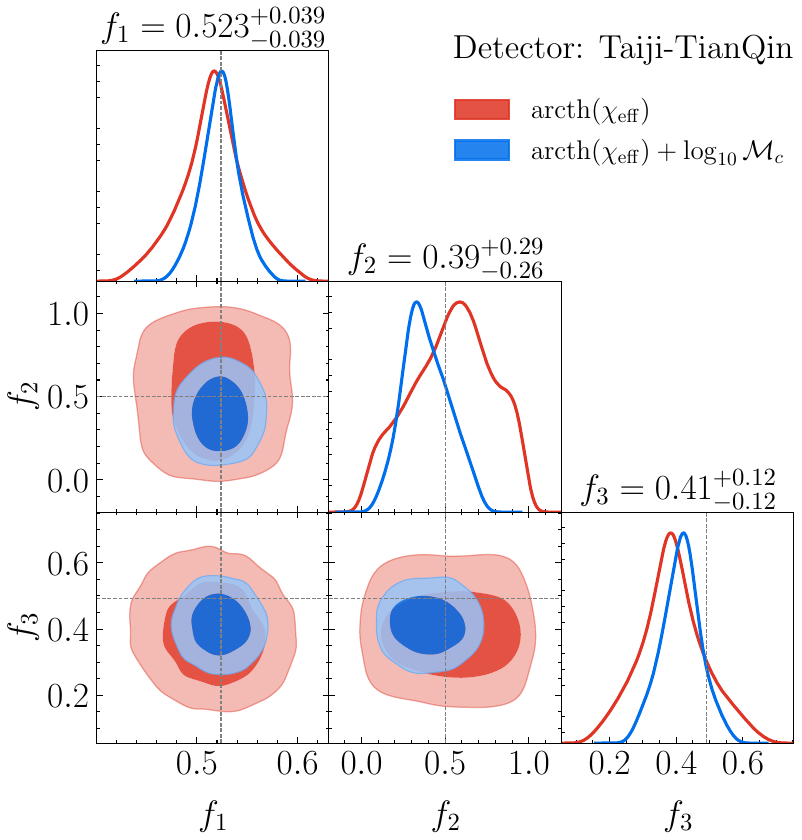}
    \caption{Posterior distributions of channel fractions $\{f_1, f_2, f_3\}$ across inference stages.
    The true values are $f_1=132/252  \approx 0.524 $, $f_2=66/132=0.5$, and $f_3=59/120  \approx 0.492$, which are marked by black lines.  The contours show the $2\sigma$ credible intervals. Step 1 (red curves) uses only $\rm{arcth}(\chi_{\rm eff})$ constraints, while Step 2 (blue curves) incorporates additional $\log_{10}\mathcal{M}_c$ information, demonstrating uncertainty reduction. 
    }
    \label{fig:channel_factor}
\end{figure}

Figure~\ref{fig:channel_factor} shows the hierarchical Bayesian inference results of the formation channel fractions, with confidence intervals progressively narrowing across the analysis stages. All true fraction values—$f_1^{\rm true} \approx 0.524$ (LS vs. HS), $f_2^{\rm true} = 0.5$ (LS\_d vs. LS\_nod), and $f_3^{\rm true} \approx 0.492$ (HS\_d vs. HS\_nod)—fall within their respective 2$\sigma$ credible intervals, confirming the validity of the inference framework.
The network effectively distinguishes between LS and HS models, achieving a relative uncertainty of 7.4\% at the $2\sigma$ level, and reliably discriminates delayed from prompt mergers within the HS channel, with an uncertainty of 24\%. The comparatively lower accuracy in classifying delay scenarios for LS systems, exhibiting a relative uncertainty of 58\%, is primarily due to greater population overlap between the LS\_d and LS\_nod scenarios.

\begin{figure}[H]
    \centering
    \includegraphics[width=0.42\textwidth]{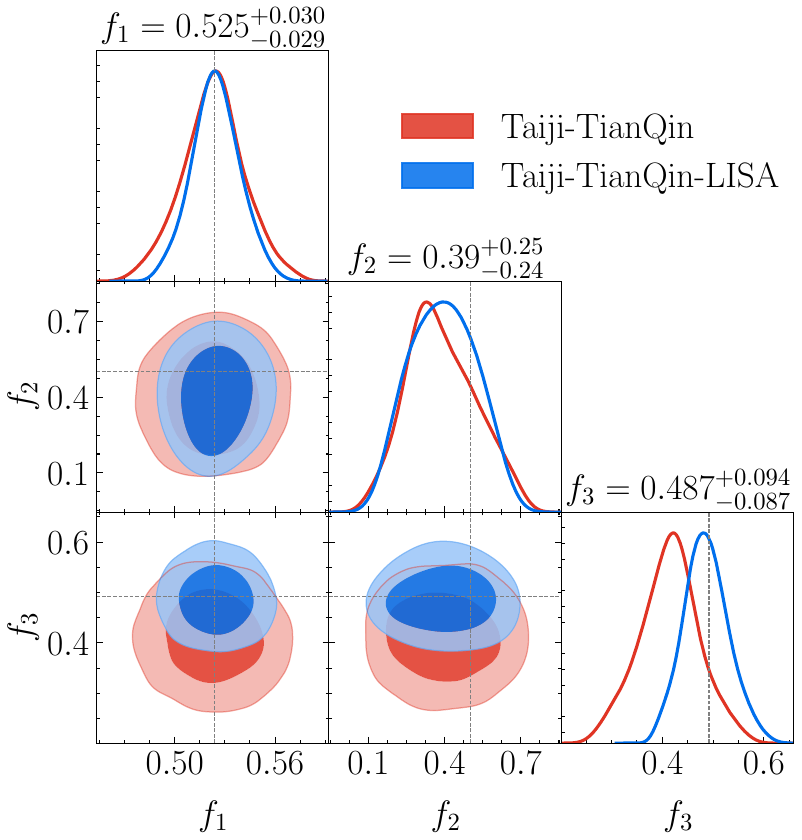}
    \caption{Posterior distributions of the channel fractions $\{f_1, f_2, f_3\}$ under different detector network configurations. The results for the Taiji-TianQin and Taiji-TianQin-LISA networks are represented by the red and blue curves, respectively.  True values $f_1^{\text{true}} \approx 0.524$, $f_2^{\text{true}} = 0.5$, and $f_3^{\text{true}} \approx 0.492$ are indicated by the vertical black lines. The contours show the $2\sigma$ credible intervals.
    }
    \label{fig:channel_factor_com}
\end{figure} 

As previously emphasized, the Chinese Taiji-TianQin network is expected to play a vital role in distinguishing between different seeding scenarios. To quantify the additional benefit of expanded detector coverage, we investigate the performance enhancement achieved by incorporating LISA to form the three-detector Taiji-TianQin-LISA network. Relative to the dual-detector Taiji-TianQin configuration, the Taiji-TianQin-LISA network improves the SNR by a factor of $1.19$--$1.25$ and increases the detection rate of MBHBs originating from LS\_{\rm d} (LS\_{\rm nod}) by a factor of $2.8$ ($3.7$). For HS models, the Taiji-TianQin-LISA network will capture more than 98\% of merger events.

Based on the larger population of detectable MBHBs,  we analyze a sample containing twice as many events for the Taiji-TianQin-LISA network as for the Taiji-TianQin configuration to infer the formation channel fractions. Fig.~\ref{fig:channel_factor_com} clearly shows that the Taiji-TianQin-LISA network yields more accurate and precise confidence intervals for all channel fraction parameters, with particularly remarkable improvement in constraining $f_3$. Specifically, it constrains the formation channel fractions with distinct precisions: $5.7\%$ relative uncertainty at the $2\sigma$ level for the LS fraction ($f_1$), $50\%$ for the delayed fraction in the LS scenario ($f_2$), and $18\%$ for the delayed fraction in the HS scenario ($f_3$). 
It provides significantly enhanced capability to distinguish between delayed and prompt mergers within the HS channel. The classification of delayed mergers in LS systems remains challenging due to the substantial population overlap between  LS\_{\rm d} and LS\_{\rm nod} channels.

\section{Conclusion}\label{sec:5}

The origin of SMBHs remains a profound mystery in astrophysics. The recent detection of a black hole at $z = 10.6$ in the galaxy GN-z11~\cite{maiolino2024smallvigorousblackhole} has revitalized interest in BH seeds. These seeds may have formed from the remnants of the PopIII stars, or could result from the direct collapse of primordial protogalactic disks in the early Universe. Unraveling the origins of SMBHs lies at the intersection of fundamental physics, cosmology, and astrophysics, bridging GW astronomy with traditional electromagnetic observations.

In this study, we conduct a detailed analysis of the Taiji–TianQin network and its potential to distinguish between different seeding scenarios. Our results demonstrate the network’s potential across three key areas: detection, parameter estimation, and the statistical discrimination of astrophysical formation channels.
First, the combined Taiji–TianQin network significantly enhances the SNR for MBHB merger signals at high redshift ($z\gtrsim 10$), improving the network SNR by a factor of 2.2–3.0 (1.06–1.14) relative to TianQin (Taiji) alone.
This improvement enables the detection of faint signals that are undetectable by individual instruments, significantly increasing the number of observable events. In particular,
the combined network greatly enhances the detection of MBHB mergers: for LS models, the detection rate increases by a factor of $2.2$ to $3$  compared to Taiji alone, while TianQin alone cannot detect such events. For HS models, the network achieves a near-complete detection efficiency, capturing more than 96\% of merger events.

Second, our Bayesian inference analysis reveals high parameter estimation precision across seven key physical parameters characterizing MBHBs. The network achieves typical relative uncertainties at the $2\sigma$ level, ranging from $10^{-4}$ for the component BH mass to $10^{-2}$ for spin-orbit tilt angles. These high precisions in measuring the component masses, spins, orientations, and luminosity distance will enable detailed characterization of MBHB systems and their origins.

Finally, the Taiji–TianQin network’s enhanced SNR and parameter estimation precision collectively support the construction of a sufficiently large and well-characterized population sample. This enables robust statistical inference of the fractional contributions of different formation channels. Using a hierarchical Bayesian framework, we sequentially incorporate $\rm{arcth}(\chi_{\rm eff})$ and $\log_{10}\mathcal{M}_c$ to constrain the formation channel fractions with distinct precisions: $7.4\%$ relative uncertainty at the $2\sigma$ level for the LS fraction ($f_1$), $58\%$ for the delayed fraction in the LS scenario ($f_2$), and $24\%$ for the delayed fraction in the HS scenario ($f_3$). The results show that the network achieves high precision in distinguishing between LS and HS origins, offers moderate discrimination for delay versus no-delay channels in HS origin binaries, and exhibits limited performance in classifying delay scenarios for LS systems—primarily due to significant population overlap between the LS\_d and LS\_nod scenarios.

Our findings are based on formation scenarios modeled within a merger-tree framework. However, it is important to note that alternative formation channels may also contribute to BH growth, such as hierarchical BH mergers occurring in active galactic nucleus (AGN) disks~\cite{Chen_2024,li2025alignedhierarchicalblackhole,Yang_2019,Rowan_2023}. As shown in Ref~\cite{Xue_2025}, the maximum BH mass from such hierarchical mergers is limited by the lifetime of AGN disks. For typical AGN lifetimes of $10$–$100~\mathrm{Myr}$, the maximum BH mass increases with lifetime but saturates near $\sim 10^3~M_\odot$, which is significantly less massive than a central SMBH. Therefore, such  merger products in AGN disks can not be seeds for  high-redshift SMBHs.   
Nevertheless, these hierarchical BH mergers could potentially interfere with our detections. We will analyze this effect in the future once the relevant binary black hole catalogs are established.

Furthermore, the James Webb Space Telescope (JWST) has revealed a new class of high-redshift ($4\lesssim z \lesssim 11$) sources, dubbed ``little red dots'' (LRDs)~\cite{jones2025mrmbhmrelationship3z7,Lin_2025,casey2024dustlittlereddots}, identified by their red rest-frame optical emission and compact morphologies~\cite{casey2024dustlittlereddots}. These objects have sparked considerable debate regarding their interpretation and physical origin~\cite{Cenci2025,zhang2025littlereddotssmallscale,akins2024cosmosweboverabundancephysicalnature}.
It remains uncertain whether LRDs can be fully explained as AGNs or whether they also represent a substantial population of massive/compact galaxies in the early Universe~\cite{akins2024cosmosweboverabundancephysicalnature}. Resolving the nature of these sources—particularly if they are reddened AGN—carries important implications for understanding black hole growth and galaxy assembly at cosmic dawn.

In the absence of definitive conclusions on early massive BH formation via the astrophysical channels discussed above, it is worth considering other, more exotic pathways~\cite{Volonteri_2003}. Upcoming GW detectors such as Taiji and TianQin will be capable of detecting increasingly distant and massive BHs, which would likely originate from exotic conditions. These include primordial BHs formed shortly after the Big Bang~\cite{carr2025primordialblackholes,zhang2025littlereddotssmallscale, Dolgov_PBH_2018}, supermassive stars sustained by dark-matter annihilation~\cite{Freese_2016}, efficient energy dissipation of magnetic fields in atomic cooling halos~\cite{2019MNRAS484185P,2009ApJ7031096S}, and fueling BHs by self-interacting dark matter~\cite{2001ApJ558482B,sabarish2025accretionselfinteractingdarkmatter}.

This work represents the first step of our group toward unveiling the origin of SMBHs. In future studies, we will use the Millennium-II~\cite{Boylan_Kolchin_2009_MRII} merger trees to model the cosmological evolution of MBHBs, incorporating additional parameters relevant to GW observations, such as eccentricity evolution and orbital inclination. 
By tracking the formation pathways of SMBHs and extracting the binary merger events along these pathways, we aim to reconstruct their assembly history and generate mock catalogs of GW signals up to very high redshift ($z \gtrsim 20$)~\cite{2023MNRAS.518.4672S}. These catalogs will allow us to evaluate the capability of space-borne GW detectors to directly observe such high-redshift BH mergers and probe the earliest phases of SMBH formation and growth.

\Acknowledgements{This work was supported by the National Key R$\&$D Program of China (Grant No. 2021YFC2203002), the National Natural Science Foundation of China (Grant No. 12473075, 12173071). This work made use of the High Performance Computing Resource in the Core Facility for Advanced Research Computing at Shanghai Astronomical Observatory.}

\InterestConflict{The authors declare that they have no conflict of interest.}

\bibliography{ref}
\bibliographystyle{scpma}

\end{multicols}
\end{document}